Title: A strategy to identify event specific hospitalizations in large health claims database

Authors: Joshua Lambert[1], Harpal Sandhu[2], Emily Kean[1], Teenu Xavier[1], Aviv Brokman[3], Zachary Steckler[3], Lee Park[3], Arnold Stromberg[3]

[1] University of Cincinnati, College of Nursing

[2] Department of Bioengineering, University of Louisville Speed School of Engineering

[3] Dr. Bing Zhang Department of Statistics, University of Kentucky

Abstract: Health insurance claims data offer a unique opportunity to study disease distribution on a large scale. Challenges arise in the process of accurately analyzing these raw data. One important challenge to overcome is the accurate classification of study outcomes. For example, using claims data, there is no clear way of classifying hospitalizations due to a specific event. This is because of the inherent disjointedness and lack of context that typically come with raw claims data. In this paper, we propose a framework for classifying hospitalizations due to a specific event. We then test this framework in a health insurance claims database with approximately 4 million US adults who tested positive with COVID-19 between March and December 2020. Our claims specific COVID-19 related hospitalizations proportion is then compared to nationally reported rates from the Centers for Disease Control by age and sex.

Funding: R21 National Library of Medicine [7R21LM013683-02](7R21LM013683-02)

Keywords: claims data, classification, COVID-19, methodology

Introduction

Prescription and health insurance claims providers can deliver unique patient-level retail pharmacy, diagnosis, and procedure data. These data can range in size and complexity depending on the provider(1–3). Successfully using these data in medical research is not an easy task and requires some key considerations(1). Some of this difficulty comes from the lack of structure and context to how certain ICD-10, current procedural terminology (CPT), or drug codes are grouped with one another around an event of interest. For example, a hospitalization CPT code does not link to the diagnosis code that caused it, nor does the drug code that was prescribed because of the event. This disjointness is a major hurdle for researchers hoping to harness these large claims data for their research question of interest.

*Motivation*

This issue arose in our own research, where we sought to use a large healthcare claims database. Like many others, Symphony Health Solutions provides its users with the data it gets from retail pharmacies, and medical claims. We then sought to use these data to reconstruct whether a patient who tested positive for COVID-19 was admitted to the hospital because of that diagnosis. We found that researchers like us (using Symphony Health Data), overcame this hurdle in a variety of ways.

*Literature Review*

To uncover how researchers deal with this lack of structure in claims data, a comprehensive literature search was conducted by a health sciences librarian (E. K.). EBSCOhost Academic Search Complete, Business Source Complete, CINAHL Plus with Full Text, MEDLINE with Full Text, and OmniFile Full Text Mega (H.W. Wilson) were searched from the dates of inception through August 2021. Additionally, the search consisted of a combination of keywords and equivalent subject headings representing "Symphony Health Solutions" as a company or a reference to the use of Symphony data. The results from the Symphony search were combined with a broad variety of terms representing the concepts of classification or categorization. An English language limit was applied to results, and after deduplication, 52 articles were retrieved.

Of the 52 articles retrieved, 32 articles were deemed relevant for this study. Of the 32 remaining articles, a clear pattern was identified as to how the authors chose to reconstruct the events of interest. Eighteen studies analyzed Symphony data using one timeline(4–21) to reconstruct the event. For example, Hampp et al.(7) used the Symphony Health Solutions PHAST Prescription Monthly database to investigate the antidiabetic drug use in the US population during a predefined single timeline. Six studies looked at data with one timeline but multiple follow-ups within a time range(22–27). Multiple timelines were used by eight of the retrieved papers (2,28–34). For example, Brixner et al.(2) conducted a longitudinal study using patient-level Symphony Health Solutions administrative claims data to assess the effectiveness of the HUMIRA Complete PSP in patients receiving adalimumab (ADA) treatment for broad range of diagnoses. They required the patients to have >2 claims which were at least 30 days apart to be included in the study.

Using this past work, as well as our own personal experience with claims data, we developed a generic methodology to reconstruct event specific hospitalizations. This proposed methodology is meant to act as a guide for how researchers can utilize health claims data in a more rigorous way.

Methods

*Event Reconstruction Strategy*

Our event reconstruction strategy centers on the overlap of various event horizons (timelines) of interests. Specifically, when it comes to identifying event specific hospitalizations, the hospitalization event horizon is an important one to define. In Figure 1, the hospitalization horizon is defined as $H^-$ to $H^+$ where each endpoint is an integer ($\mathbb{Z}$). Event specific hospitalizations, by definition, must have the event of interest occur within the hospitalization horizon ($H^-$, $H^+$). Other relevant/conditional event horizons may be defined ($O^-$ to $O^+$ or $A^-$ to $A^+$) to sensitize the definition around the time of the hospitalization. The other event horizon(s) may act as validation event(s) which may come from a contextual understanding of the problem or a literature search. If a specific patient has the event of interest within the hospitalization horizon and, if necessary, the other relevant/conditional event horizons occur within the designated horizon around the hospitalization then the patient is said to have had the event specific hospitalization. All other patients can be thought of as not having the event. In Figure 1 (as represented by stars) an example patient had a hospitalization 2 days after the event of interest and had one conditional event 3 days after their hospitalization and another conditional event 2 days after their hospitalization. Because this patient had events within the designated horizons ($H^-$ to $H^+$, $O^-$ to $O^+$ and $A^-$ to $A^+$) they are said to have the event specific hospitalization.

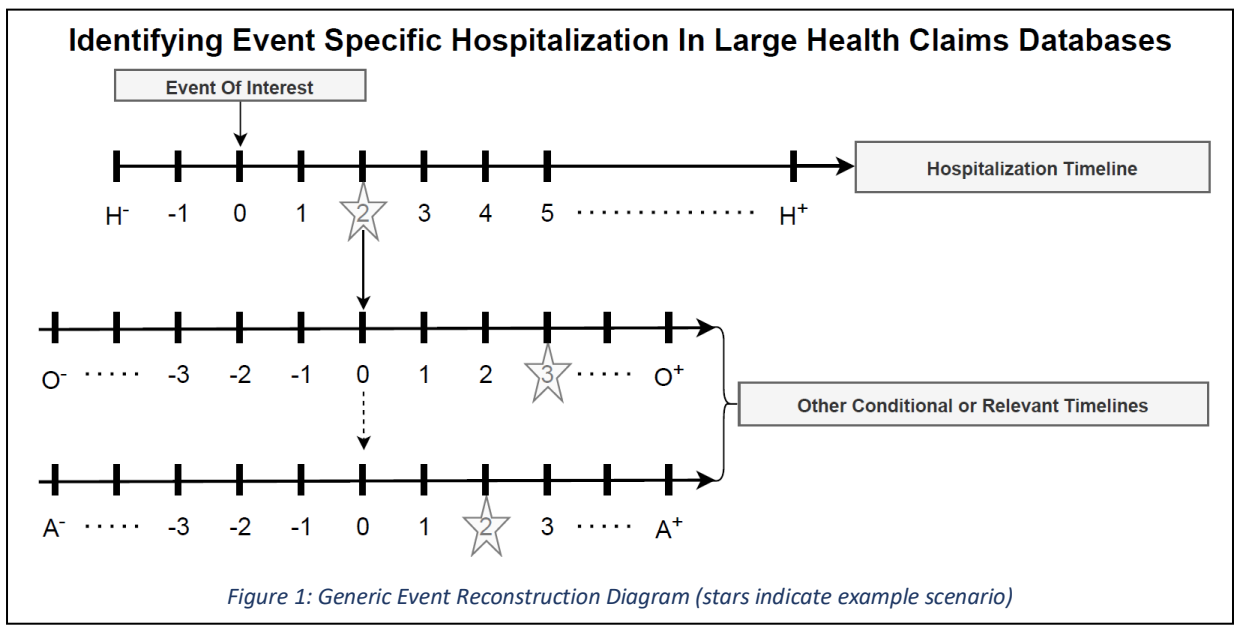

*Figure 1: Generic Event Reconstruction Diagram (stars indicate example scenario)*

Results

The COVID-19 research database enables public health and policy researchers to use real-world data to better understand and combat the COVID-19 pandemic. In June 2021, via the COVID-19 research database, we gained access to the Symphony Health Data. Our symphony data had approximately 4 million patients who tested positive for COVID-19 between 03/01/2020 and 12/31/2020. While we had data after 12/31/2020, our study focused on 2020 due to access to the FDA emergency use authorization (EUA) COVID-19 vaccines in early 2021 and beyond which we felt would deduct from our focused study of interest. Patient records are de-identified and minimal demographic information (Age, Sex, first two digits of the patient's residential zip code) is known about the unique patients within the dataset. Patient level CPT, diagnosis, and prescription codes were available from late 2018 to mid-2021. Using these data, which are contained in different tables accessible via the snowflake SQL platform, we utilized our event reconstruction strategy where we intended to reconstruct which, of the 4 million patients who tested positive for COVID-19, were hospitalized due to the COVID-19 diagnosis.

*Hospitalization Due To COVID-19 Diagnosis Reconstruction*

Using our generic definition, defined above, and outlined in Figure 1, our clinical and research team decided the necessary event horizons endpoints. First, the hospitalization CPT code (99221, 99222, or 99223) needed to occur in the -2 to 14-day timeline from COVID diagnosis (U07.1). This time lag and lead were determined as claims do not always mimic the actual timeline that the patient experienced. A sensitivity analysis of these endpoints showed that most of the diagnoses and hospitalizations that met our -2 to +14 criteria actually occurred very close (-1 to +1) to one another. If multiple hospitalization CPT codes or multiple COVID-19 diagnoses codes occurred for a specific patient, then the minimum distance between all possible combinations of diagnoses and hospitalizations were considered. As a tie breaker the earliest minimum combination which met our criteria was considered as the COVID-19 hospitalization for that specific patient. If one of the combinations met the criteria of -2 to +14 then the patient was said to have met the first part of the criteria for being classified as hospitalized due to a COVID-19 diagnosis. As a validation, patients needed to have at least one of a set of additional diagnoses which occur around the time (-14 to +7) of a hospitalization. This set was again, determined by our clinical and research team. These were: pneumonia due to SARS-associated coronavirus (J12.81), other viral pneumonia (J12.89), acute bronchitis due to other specified organism (J20.8), bronchitis not specified as acute or chronic (J40), unspecified acute lower respiratory infection (J22), other specified respiratory disorders (J98.9), or acute respiratory distress syndrome (J80). For example (See Figure 2), if a patient had a COVID-19 diagnosis, 2 days before they were hospitalized, and had one of the other additional diagnoses 3 days after the hospitalization then that patient would be defined as a patient who was hospitalized due to a COVID-19 diagnosis.

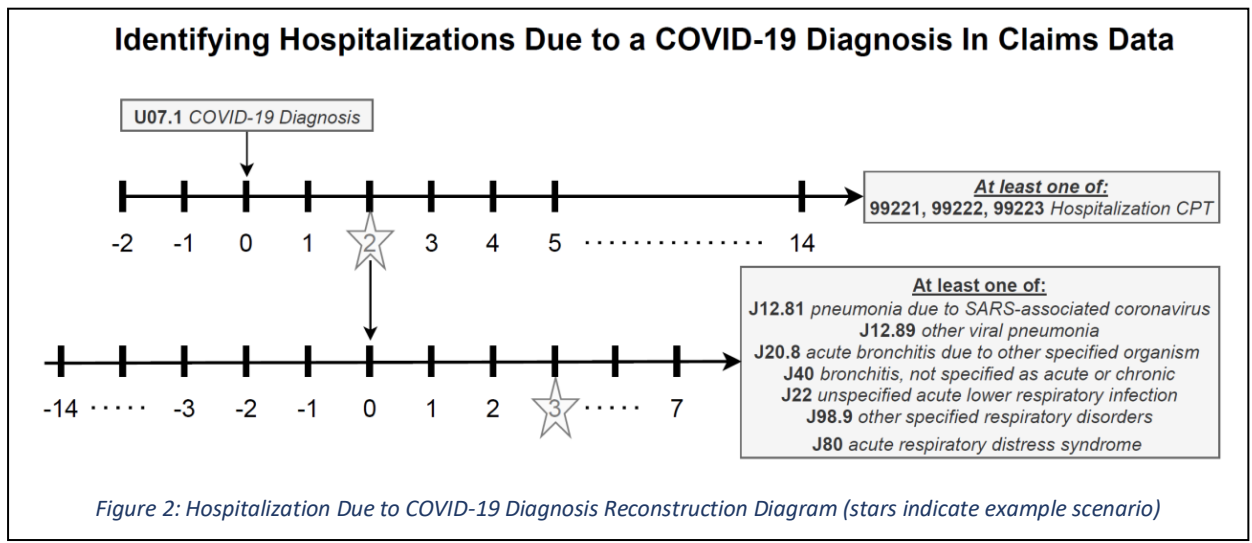

Figure 2: Hospitalization Due to COVID-19 Diagnosis Reconstruction Diagram (stars indicate example scenario)

*Comparison Of Proposed Methodology and CDC Estimates*

The CDC provides estimates(35) of the number of symptomatic COVID-19 illness and the aggregate number of hospitalizations from February 2020 through May 2021. Unfortunately, we were unable to attain estimates for the same time period under investigation in our study (March 1st, 2020, through December 31st, 2020). Our methodology was developed independently of the CDC data, and the CDC data is meant to provide a type of external validation to our methodology.

Table 1 provides a comparison of how our methodology on the Symphony data compares to the CDC's population estimates. Using the proposed methodology on the Symphony data, 2.7% of the patients in the age group of 18-49 years old were hospitalized due to a COVID-19 diagnosis. The CDC's estimated that 3% of 18-49 years old were hospitalized due to symptomatic COVID-19. In the age group of 50-64 years old, 8.2% of the Symphony patients were hospitalized as compared to the 9.2% estimated by the CDC. In the age group of 65+ years old, 14.6% of the Symphony patients were hospitalized as compared to the 28.1% estimated by the CDC. Across all age groups the total percentage of Symphony patients hospitalized due to COVID-19 was 7.3% which is similar to the estimate of 7.5% by the CDC.

The difference observed in the 65+ population is worrisome and warrants serious attention. Upon reinvestigation these results were not surprising as our Symphony data are from private insurance claims data and their estimates would be for those United States adults who were privately insured. In the United States, those who are 65+ years old qualify for Medicare and may not be privately insured. One explanation as to why so few 65+ in the Symphony data are not hospitalized due to a COVID-19 diagnoses (as compared to the CDC's estimates) is that they may be wealthier (can afford private insurance) and are therefore healthier on average.

| Age Groups | CDC Estimated COVID-19 Cases | Symphony Estimated COVID-19 Cases | CDC Estimated Hospitalizations | Symphony Estimated COVID-19 Hospitalizations | CDC Estimated Percentage of Hospitalizations (*) | Symphony Estimated Percentage of Hospitalizations (#) |
|---|---|---|---|---|---|---|
| 18-49 years | 51,581,445 | 1,867,749 | 1,533,679 | 49,638 | 3.0% | 2.7% |
| 50-64 years | 17,377,602 | 1,045,133 | 1,604,612 | 85,271 | 9.2% | 8.2% |
| 65+ years | 10,005,696 | 1,061,390 | 2,808,089 | 155,439 | 28.1% | 14.6% |
| All ages | 78,964,743 | 3,974,272 | 5,946,380 | 290,348 | 7.5% | 7.3% |

\* CDC Estimated Percentage of Cases= CDC Estimated COVID-19 Hospitalizations/CDC Estimated COVID-19 Cases
\# Symphony Estimated Percentage of Hospitalizations=Symphony Estimated Hospitalizations/Symphony Estimated COVID-19 Cases

*Table 1*

While our sample is older on average than the general population, our methodology's overall estimates for the percent of COVID-19 diagnoses who were hospitalized due to a COVID-19 diagnosis is close to the CDC's overall estimates (7.3% vs. 7.5%). The main place to compare our methodology to the CDC estimates is in the portion of the US adult population who is likely privately insured (18-64 years). Within this group, our methodology seems to capture a percent of the sample similar to that of the CDC estimates.

Conclusion

A strength is that the methodology is a rigorous way to define cases in claims data based on multiple time horizons. In addition, this methodology can be used for other diseases events as well as other claims databases. Our attempt to validate this methodology vs the CDC's estimates showed similar estimates within those likely to be privately insured (18-64). The Symphony data has a high coverage rate: 92% of retail pharmacy claims, 71% of mail-order pharmacy claims, and 65% of specialty pharmacy claims, according to COVID-19 Research Database(36). No method is without its fair share of real limitations and considerations. First, no methodology will be able to take raw claims and produce results free from disputes as there will always be a lack of context and structure. Defining the time horizons is difficult, as there is likely great differences